# Safer Traffic Recovery from the Pandemic in London
## -- Spatiotemporal Data Mining of Car Crashes

Kejiang Qian, Yijing Li[*]

CUSP London, Department of Informatics, King's College London, London, UK.

**Abstract**

In the aim to support London's safer recovery from the pandemic by improving road safety intelligently, this study investigated the spatiotemporal patterns of age-involved car crashes and affecting factors, upon answering two main research questions: (1) "What are the spatial and temporal patterns of car crashes as well as their changes in two typical years, 2019 and 2020, in London, and how the influential factors work?"; (2) "What are the spatiotemporal patterns of casualty by age groups, and how people's daily activities affect the patterns pre- and para- the pandemic"? Three approaches, i.e., spatial analysis (network Kernel Density Estimation, NetKDE), factor analysis, and spatiotemporal data mining (tensor decomposition), had been implemented to identify the temporal patterns of car crashes on weekly and daily basis respectively, detect the crashes' hot spots, and to gain better understanding the effect from citizens' daily activity on crashes' patterns pre- and para- the pandemic. It had been found from the study that car crashes mainly clustered in the central part of London, especially busier areas around denser hubs of point-of-interest (POIs); the POIs, as a reflector for citizens' daily activities and travel behaviours, can be of help to gain a better understanding of the crashes' patterns, upon further assessment on interactions through the geographical detector; the crashes' casualty patterns varied by age group, with distinctive relationships between POIs and crashes' pattern for corresponding age group categorised. In all, the paper provided an in-depth exploratory analysis of car crashes and their casualty patterns in London to facilitate deployment policies towards post-pandemic safer recovery upon COVID-19.

---

[*] Corresponding author.
*E-mail address:* kejiang.qian@kcl.ac.uk (K. Qian), yijing.li@kcl.ac.uk (Y. Li)

Declarations of interest: none

This research did not receive any specific grant from funding agencies in the public, commercial, or not-for-profit sectors.

# 1 Introduction

Car crashes have long been recognised as exerting tremendous effects on economic development and human well-beings among societies. Most people involved in road traffic crashes will get injured or even die from the accidents, where the number of crash deaths peaked at 1.35 million around the world in 2016, with more than half of fatalities aged between 15 and 44 (World Health Organization (WHO, 2018). London also engenders such challenges and has around 80,000 reported collisions during 2017-2019 before the pandemic in 2020, when people aged between 25 and 59 accounting for the most casualties, whilst those over 60 have a higher fatality rate. Such an issue was predicted to get worse alongside the rapid population growth, as the Greater London Authority (GLA) projected a potential population of around 11 million in 2050 based on 8.17 million citizens censused in 2011 (GLA, 2020). On the other hand, the COVID-19 pandemic exerted significant influences on road traffic activities with the figure that, UK Office for National Statistics (ONS, 2020) reported up to 40% of workers work from home during the pandemic, dramatically reducing traffic demand and expected to have a long-term effect on people's daily travel behaviour (Katrakazas et al., 2020) and transport mode preference (Batty, 2020). For example, the public gradually lost confidence in public transport but instead increased private car usage (Eisenmann et al., 2021), which is expected to result in more traffic congestion and collisions upon recovery from the pandemic. Therefore, studies on road safety by varied daily patterns among different age groups were considered to be in a better position formulating car crash prevention deployment strategies.

Empirical studies on crash distributions identified the spatial and temporal variations that, crashes tend to occur more around working places rather than the residential places (Levine et al., 1995), and had been thought to be affected by multiple factors, e.g., daily human activities and build environment factors (Ziakopoulos and Yannis, 2020), requiring for further investigation into quantified proxy of point-of-interest (POI) data (Worachairungreung et al., 2021). The POIs hereby represent functional places' geographical locations, including restaurants, shops, schools, parks, etc., and could reflect the injury patterns affected by daily human activities by different age groups. For instance, Jha et al. (2003) had identified the highest casualties from car crashes fell into the age group 20-29. However, there were still few research examining the interactive effect between multiple POIs and traffic crashes by age groups, making the topic worthy of further exploration.

Being motivated by the incentive to improve road safety intelligently and support traffic recovery, this study aims to identify spatiotemporal patterns of age-categorised traffic crashes, and further to investigate the influential factors towards severer crashes pre- and para- pandemic. Besides, to gain richer knowledge of traffic crashes' variations among age groups in London pre- and para- pandemic, two main objectives are expected to be arrived at:

- The spatiotemporal pattern and varied recognition of car crashes in two different representative years, and the underlying influences from POI factors.
- The spatiotemporal casualty patterns by age group, as well as the effects from people's varied daily activities pre- and para- the pandemic.

To realise such objectives, this study will (1) analyse the spatiotemporal characteristics of car crashes by the severity; (2) construct the geographical detector quantifying individual and interactive effects of POI categories on crash distribution; (3) extract the spatial and temporal crashes patterns of victims by age groups; and (4) measure the relationship between their daily activities and corresponding patterns of car crashes by influences of POIs categories based on the road safety data in 2019 and 2020, respectively.

## 2 Literature review

A large body of literature on traffic crashes pattern identification had been realised through spatial analysis and factor analysis. The spatial analysis techniques include examining car crashes' first-order effect, i.e., the hot spot of car crashes detection, and the second-order effect, i.e., spatial clusters detection, with Kernel Density Estimation (KDE) as a widely adopted approach. Although there are several studies utilizing planar KDE to map traffic crashes (Blazquez and Celis, 2013), biased estimations of crashes' density values over 2-D Euclidean space had been spotted since the crashes mainly accumulated along with the road networks (Yamada and Thill, 2007). It led to Okabe et al.'s proposing network-based KDE (NetKDE) estimating crashes along the roadway and the development of the SANET toolbox, followed by an increasing number of studies on car crashes using the NetKDE approach (Xie and Yan, 2008, Zahran et al., 2021). Besides, Koloushani et al. (2022) found that hot spots of young-driver-involved crashes data supported the justification that NetKDE has a better performance than planar KDE when receiving localized focuses. The second-order effect had traditionally been measured by widely recognised Moran's *I* index, in the aim to estimate the strength of spatial autocorrelation and discover statistically significant risk clusters. For example, Iyanda (2019)'s work explores the distribution of accident severity, and Shabanikiya et al. (2020)'s work identifies high-risk areas of pedestrian crashes with children's involvement.

To investigate the influences on traffic crashes from environmental factors, factor analysis had been utilised as the mainstream with car crashes being treated as point events, and the count data models as the basic methods embracing spatial heterogeneity existence. For example, the spatial zero-inflated negative binomial model (Champahom et al., 2020), and the fixed bandwidth geographically and temporally weighted ordered logistic regression model (Chen et al., 2022). However, these models failed to capture the dynamic interactions between selected factors and traffic crashes. To fulfill such a gap, geographical detector proposed by Wang et al. (2016) had been considered in light of its being excellent at measuring the influence of factors on geographical phenomena based on spatial stratified heterogeneity analysis, and received extensive applications in varied research fields, like housing price (Wang et al., 2017) and air pollution (Ding et al., 2019), as well as car crash analysis (Zhang et al., 2020) assessing injury factors' influences on casualties considering their mutual interaction, but still leaving the research on interactions of factors for road safety less explored.

In parallel, spatiotemporal data mining technique has been taken as a more efficient method exploring patterns of spatiotemporal data (Han et al., 2012), in comparison with traditional spatial analysis methods. This data mining rooted method on spatiotemporal data mining and urban computing (Zheng et al., 2014a) was represented by the most favourite approach, tensor decomposition (Kolda and Bader, 2009). Tensor decomposition has been advantaged by its modelling multidimensional data and dealing with data sparsity issues, with example applications in analysing human mobility from OD-matrix data and decomposing the spatiotemporal patterns of human mobility through measuring the variability of multidimensional mobility patterns (Yao et al., 2015, Sun and Axhausen, 2016); Zheng et al. (2014b) also applied it onto New York City's sparse noise data analysis, and tried to recover the noise level by combining with POIs, road network, and social media data. However, there was still very limited literature applying it in traffic crash research.

In a nutshell, the limitations of empirical studies (summarized in Table 1) rendered this study the opportunity for improvements in: combining spatiotemporal data mining with spatial analysis and factor analysis techniques, onto age-categorised traffic crashes analysis from three dimensions which are time, space, and casualty severity by age group, to highlight the age-bounded effects from traffic crashes.

Table 1 Limitations of current studies

| Authors & Year | Regions | Methodologies | Limitations |
|---|---|---|---|
| Okabe et al. (2006); Blazquez and Celis (2013); Koloushani et al. (2022) | Tokyo, Japan Santiago, Chile Florida, USA | Planar KDE & NetKDE | This method has worse performance than NetKDE when there is a localized focus. |
| Wang et al. (2016); Ding et al. (2019); Zhang et al. (2020) | China; China; Shenzhen, China | Geographical detector | There are rare studies of crash factor analysis by considering specific risky indicators by age breakdown, the relationship of the POIs, |

| | | | | |
|---|---|---|---|
| Zheng et al. (2014b); Yao et al. (2015); Sun and Axhausen (2016) | New York, USA; Massachusetts, USA; Singapore | Tensor decomposition | and the interactive effects of factors. |
| | | | This method is very useful, but there was only very limited literature applying it in traffic crash research. |

# 3 Data and Methodology

London, as the capital of the United Kingdom, has been chosen as the study area with its 4,835 Lower Super Output Area (LSOA) units (GLA, 2014), where each has an average population of 1500 (Green et al., 2011).

## 3.1 Data acquisition

Road safety data of London had been collected from the GOV.UK in point format at LSOA scale from 01 January to 30 December 2019 and from 01 January to 29 December 2020, with detailed information on recorded car crashes at 29,023 in 2019 and 23,551 in 2020 respectively (Table 2). Among the records, 84.47% in 2019 and 85.3% in 2020 were slight, but 0.57% in 2019 and 0.53% in 2020 were fatal. In reference to Department for Transport (DfT)'s 11 breakdown categories of the casualties by age group, this study restructured the age groups into seven research conveniences as listed in Table 2 and Table 3.

Table 2 Car crashes and causalities by age group (2020 vs. 2019)

| Age groups | Frequency of slight crashes | | Frequency of serious crashes | | Frequency of fatal crashes | | Frequency of all crashes | |
|---|---|---|---|---|---|---|---|---|
| | 2020 | 2019 | 2020 | 2019 | 2020 | 2019 | 2020 | 2019 |
| 0–18 | 2108 | 2858 | 376 | 498 | 5 | 14 | 2489 | 3370 |
| 19–25 | 3525 | 3886 | 523 | 730 | 17 | 29 | 4065 | 4645 |
| 26–35 | 6087 | 6846 | 845 | 1048 | 34 | 34 | 6966 | 7928 |
| 36–45 | 3845 | 4786 | 618 | 770 | 22 | 21 | 4485 | 5577 |
| 46–55 | 2544 | 3331 | 489 | 588 | 17 | 16 | 3050 | 3935 |
| 56–65 | 1294 | 1679 | 275 | 331 | 7 | 14 | 1576 | 2024 |
| Over 65 | 686 | 1131 | 211 | 375 | 23 | 38 | 920 | 1544 |
| Total | | | | | | | 23551 | 29023 |

*Note: Fatal accident refers to an accident in which at least one person is killed; other casualties (if any) may have serious or slightly injuries; serious accident means one in which at least one person is seriously injured but no person (other than a confirmed suicide) is killed; slight accident represents one in which at least one person is slightly injured but no person is killed or seriously injured. Definitions retrieved from official document at https://assets.publishing.service.gov.uk/government/uploads/system/uploads/attachment_data/file/259012/rrcgb-quality-statement.pdf*

It could be read from Table 2 that, the car crashes and causality in 2020 had decreased from the level in 2019 during the pandemic, but with a similar distribution pattern across age group breakdowns; the frequency of all crashes or by varied severity in 2020 had been decreased proportionally by age group against 2019. Since this study is aiming to explore the pattern and mechanism of safer traffic recovery from pandemics, 2019 and 2020 as the representative normal and abnormal years, respectively, would be sufficient to suggest the patterns, hence shed light on the traffic recovery strategies upon recovery from 2020 back to normal. So the following analysis will focus on the case study in 2019 and 2020 to explore the patterns in detail.

POI datasets in this study had been collected from the Ordnance Survey, consisting of geospatial information on all London business, entertainment, education, etc. in 2019, with ten categories: accommodation, eating and drinking, attractions, commercial services, sport and entertainment, education and health, public infrastructure, transport, manufacturing and production, and retail, as summarised and represented as relevant icons in Table 4.

Table 4 Statistical summary of POI categories in London

| POI category | Count | Mean | Std | Min | Max |
| --- | --- | --- | --- | --- | --- |
|  (WORK) | 142,005 | 29.369 | 4685.475 | 0 | 3218 |
|  (ATTRACTION) | 11,269 | 2.329 | 41.347 | 0 | 207 |
|  (EDUCATION) | 11,969 | 2.475 | 8.373 | 0 | 8 |
|  (SHOP) | 49,990 | 10.339 | 578.338 | 0 | 759 |
|  (RESTAURANT) | 29,708 | 6.144 | 18.271 | 0 | 807 |
|  (STATION) | 21,209 | 4.386 | 25.098 | 0 | 140 |
|  (ENTERTAINMENT) | 12,691 | 2.624 | 4.432 | 0 | 118 |
|  (ACCOMMODATON) | 1,749 | 0.362 | 2.525 | 0 | 53 |
|  (PARKING) | 1,533 | 0.317 | 0.872 | 0 | 27 |

## 3.2 Severity-weighted index

Severity-weighted index (SWI) had been calculated to indicate the seriousness of crashes rather than purely the volume of crashes, based on the combined 5-3-1 weighting system (Geurts et al., 2004) with each fatal injury weighted at 5, serious injury weighted at 3, slight injury weighted at 1 in Equation (1):

$$SWI = 5 \times x_{fatal} + 3 \times x_{serious} + x_{slight} \tag{1}$$

where $x_{fatal}$, $x_{serious}$ and $x_{slight}$ respectively presents for the total number of fatal, serious, and slight traffic crashes.

**3.3 Spatial analysis**

Spatial analysis of crash distributions on basis of road networks and planar maps at LSOA level, to discover the first-order and second-order effects through NetKDE and Moran's *I* statistics respectively.

*3.3.1 Hot spots detection based on the NetKDE*

NetKDE targeting at point events in a network space is an extension of planar KDE with its two main specialties (Xie and Yan, 2008): (1) network distance $d_{is}$ from crash point $i$ to location $s$ is calculated as the shortest-path distance along the road network; and (2) the density estimator is computed per linear unit as defined below in Equation (2):

$$\lambda(s) = \frac{1}{r}\sum_{i=1}^{n} w_i \cdot K(\frac{d_{is}}{r}) \tag{2}$$

where $\lambda(s)$ presents the density estimator at location $s$; $r$ is the bandwidth of the KDE, and $w_i$ is the weight of crash point $i$. $K$ is the kernel modeled as the kernel function of distance $d_{is}$. The method assumes that crashes occur alongside roadways. Crashes data need to be snapped on the network within a 10-metre road network distance, because previous crashes may not be recorded on road networks due to issues such as the low precision of the measurement system or car moving right after a crash.

Xie and Yan (2008) also pointed out that the network needs to be divided into an equal-length linear unit, which may have an impact on the local variation details. So this study selects the 200-metre road segment as the basic unit considering London's vast road network, in order to decrease the number of linear units so for a good balance between computational efficiency and accuracy. On the other hand, the choice of bandwidth also affects performance of the network KDE, As suggested by (Okabe and Sugihara, 2012) an optimal bandwidth should be 100–300 metres would be optimal, so this study designed a bandwidth of 200 metres for network KDE and Gaussian function as the kernel function. It is hypothesised that network KDE can improve the drawback of planar KDE (Xie and Yan, 2008) on over-detecting clustered point events, hence enhancing the accuracy of density estimation in intersections.

### *3.3.2 Spatial clustering identification using Moran's I statistics*

Spatial autocorrelation can reflect the strength of the spatial dependence between factors, so Moran's *I* statistics could be utilised to identify clusters with similar SWIs or the dispersion of those with dissimilar SWIs. The global Moran's *I* statistic, denoted as $I$, can be computed as in Equation (3):

$$I = \frac{n \sum_{i=1}^{n} \sum_{j=1}^{n} w_{ij} d_i d_j}{\sum_{i=1}^{n} \sum_{j=1}^{n} w_{ij} \sum_{i=1}^{n} d_i^2} \qquad (3)$$

where $n$ is the total number of factors, $d_i$ and $d_j$ present the deviations of the $i$-th and $j$-th LSOA's SWI to their means: $(SWI_i - \overline{SWI})$ and $(SWI_j - \overline{SWI})$, and $w_{i,j}$ is the spatial weight between factor $i$ and $j$, which represents the relationship between research units under Queen criterion. Down to the local indicators of spatial association (LISA) measuring the degree of local spatial autocorrelation of SWIs could be expressed in Equation (4) as:

$$LISA_i = z_i \sum_j w_{ij} z_j \quad z_i = \frac{SWI_i - \overline{SWI}}{\sqrt{\frac{1}{n}\sum(SWI_i - \overline{SWI})^2}} \quad (4)$$

where LISA clusters denoted as High-High (HH) illustrate the high-valued target clustered with high-valued neighbours; the low-low (LL) cluster indicates for low values are clustered with each other; the high-low (HL) or low-high (LH) clusters are high values surrounded by low values and vice versa.

### 3.4 Geographical detector for factor analysis

Upon exploring the spatial clustering of crashes, it is necessary to quantify the influence of POI factors on traffic crashes through geographical detectors, which can reflect daily human activities for our in-depth interpretation. In this study, it is assumed that if road traffic crashes were affected by POI categories, they may follow a similar spatial distribution pattern. Among the four geographical detectors, which are factor detector, interactive detector, ecological detector, and risk detector, the former two are deployed in this study to measure the nonlinear or linear relationship between factors. The factor detector is used to find the dominant factor of the geographical phenomena, while the interactive detector is applied to measure the interconnected effects between pairs of factors on crashes' distributions.

The factor detector computes the value of the power determinant (PD) through $q$ statistics, denoting the percentage of explainability for target variable by another variable, which is similar to R-square statistics of regression models, as in Equation (5):

$$PD = 1 - \frac{1}{N\sigma^2}\sum_{h=1}^{l} N_h \sigma_h^2 \qquad (5)$$

where $N$ denotes the number of LSOA units, $\sigma^2$ and $\sigma_h^2$ represent the variance of the dependent variable of the whole study area and that in the units $h$, respectively; the PD value indicates the strength of the spatially stratified heterogeneity of crash distributions and the contribution of the POIs to their spatial pattern.

The interaction detector is good at estimating the interactions between two individual factors. If the PD value of interaction is greater than the accumulation of the PD values from individual effect, a significant enhancement influence from the interaction is present. This study will build up the model with the number of POI categories in each LSOA as an explanatory variable, and numeric variables need to be stratified for detectors; whilst Jenks's natural breaks (JNB) method is applied to determine the optimised categorisation of each POI, in the criteria of minimising the variances within classes but maximising them between classes (Chen et al., 2013).

## 3.5 Spatiotemporal pattern detection using tensor decomposition

Tensor decomposition excels at analysing the multidimensional road safety data in time and space patterns, hence being able to decide the various crash scenarios for geographical detector methods to fit in. For example, if a crash pattern involving minors during rush hours has been identified with a high PD value between POIs of education, that indicates a scenario about students & rush hours.

### *3.5.1 Tensor construction and decomposition*

An *N*th-order tensor is a multidimensional matrix of *N* vector spaces, with each vector as a first-order tensor (*N*=1) and the matrix as a second-order tensor (*N*=2), but in vision to present three- or higher-dimensional data hereby with a higher order of the tensor. For example, three-dimensional data can be written in a tensor as $\mathcal{A} \in \mathbb{R}^{I_1 \times I_2 \times I_3}$ with three dimensions denoting $I_1$, $I_2$, and $I_3$. To extract and discover latent patterns from tensors, approaches such as singular vector decomposition (SVD), latent factor modeling (LFM), and principal component analysis (PCA) are required. However, such methods may cause information loss due to corrupted data, and make tensor decomposition outstanding with much better performances (Kolda and Bader, 2009). The most efficient tensor decomposition method is tucker tensor decomposition (Tucker, 1966), to decompose the tensor $\mathcal{A} \in \mathbb{R}^{I_1 \times I_2 \times \cdots \times I_N}$ into a core tensor $\mathcal{G} \in \mathbb{R}^{J_1 \times J_2 \times \cdots \times J_N}$ and a set of nonsingular matrices $\mathbf{A}^{(1)} \in \mathbb{R}^{I_1 \times J_1}$, $\mathbf{A}^{(2)} \in \mathbb{R}^{I_2 \times J_2}$, $\cdots$, and $\mathbf{A}^{(n)} \in \mathbb{R}^{I_n \times J_n}$. It should be noted that the size of the core tensor must be defined manually, which may affect the performance of the model in Equation (6).

$$\mathcal{A} \cong [\![\mathcal{G}; \mathbf{A}^{(1)}, \mathbf{A}^{(2)}, \cdots, \mathbf{A}^{(n)}]\!] = \mathcal{G} \times_1 A^{(1)} \times_2 A^{(2)} \cdots \times_n A^{(n)} \tag{6}$$

where $\mathcal{G} \times_n \mathbf{A}$ is the n-mode tensor-matrix product of a tensor $\mathcal{G} \in \mathbb{R}^{J_1 \times J_2 \times \cdots \times J_N}$ with a matrix $\mathbf{A}^{(n)} \in \mathbb{R}^{I \times J_n}$, which means multiplying a tensor by an n-mode matrix and is defined in Equation (7):

$$[\mathcal{G} \times_n \mathbf{A}]_{j_1, \cdots, j_{n-1}, i, j_{n+1}, \cdots j_N} = \sum_{j_n=1}^{J_n} \mathcal{G}_{j_1, \cdots, j_{n-1}, i, j_{n+1}, \cdots j_N} \mathbf{A}_{i, j_n} \in \mathbb{R}^{J_1, \cdots, J_{n-1}, I, J_{n+1}, \cdots J_N} \tag{7}$$

### 3.5.2 *Nonnegative Tucker decomposition*

The Tucker tensor decomposition also has a drawback on possible negative elements in its results, which deviates from the fact that no negative SWI of crashes data, hereby calls up the utilisation of nonnegative Tucker decomposition (NTD) (Shashua and Hazan, 2005) to find a nonnegative core tensor $\mathcal{G}$ and nonnegative matrices $\mathbf{A}^{(1)}$, $\mathbf{A}^{(2)}$, $\cdots$, $\mathbf{A}^{(n)}$, and be modelled in Equation (8) for optimisation:

$$minimize \ \frac{1}{2} \left\| \mathcal{G} \times_1 A^{(1)} \times_2 A^{(2)} \cdots \times_n A^{(n)} - \mathcal{A} \right\|_F^2 \tag{8}$$

$$\text{subject to } \mathcal{G} \in \mathbb{R}^{J_1 \times J_2 \times \cdots \times J_N} \geq 0, \ \mathbf{A}^{(n)} \in \mathbb{R}^{I \times J_n} \geq 0 \ \forall n = 1, 2, 3 \cdots, N$$

where $\|\cdot\|_F$ presents the Frobenius norm of a tensor $\chi \in \mathbb{R}^{I_1 \times I_2 \times \cdots \times I_N}$ defined as:

$$\|\chi\|_F = \sqrt{\sum_{i_1=1}^{I_1} \cdots \sum_{i_N=1}^{I_N} x_{i_1, i_2, \cdots i_N}^2} \tag{9}$$

Since the size of the core tensor influences the capability of the NTD model, it is crucial to find an appropriate size on basis of Kullbask-Leibler (KL) divergence, which is also known as the relative entropy and is widely used to measure the similarity between two discrete density distributions $M$ and $N$ (Hershey and Olsen, 2007).

$$D_{KL}(M\|N) = - \sum m(x) log n(x) + \sum m(x) log m(x) = H(M,N) - H(M) \qquad (10)$$

where the entropy of distribution $M$ is referred to as $H(M)$, and the cross-entropy of distributions $M$ and $N$ is presented as $H(M,N)$. Additionally, the following equation illustrates the measurement of the original tensor $\mathcal{A} \in \mathbb{R}^{i \times j \times k}$ and decomposed tensor $\hat{\mathcal{A}} \in \mathbb{R}^{i \times j \times k}$ based on the KL divergence: the smaller value is, the closer between two distributions; and the optimal size will be indicated by the converging point of the minimum value.

## 4 Results

### 4.1 Temporal pattern of road traffic crashes

The SWI index of traffic crashes (accumulative) and their temporal patterns in 2019 and 2020 are summarized in Figure 1 (a, b, c), illustrating similar distribution patterns. The daily patterns in either time-of-day or day-of-week could be presented in Figure 1(a), where the *y-axis* indicates day-of-week, the *x-axis* stands for time-of-day by every two hours, and the legend on the right scaled values of SWI. It is obvious that crashes occur more frequently on weekdays than on weekends; the peak hours on weekdays are routinely 8 am to 9 am and 5 pm to 7 pm;

comparatively, weekends have a higher risk of traffic crashes at midnight in these two years. An exceptional observation on Wednesday afternoon with a minor peak time 3 pm to 4 pm might owe to its routinely being the sports day after 2 pm. However, there was a small difference that the maximums of the SWI during the pandemic were less than those in 2019. This phenomenon is also illustrated in Figure 1(b) & (c) on the characteristics of crashes' breakdown by seven age groups on weekdays and weekends, respectively. Besides, Figure 1(b) clearly shows the higher causality for 26–35 years old during peak hours on weekdays, followed by the relatively higher causality for children and youngsters (0–18 and 19–25) after school. Weekends' car crashes rarely happen at midnight, as reflected in Figure 1(c), but more tend to accumulate among young and middle-aged groups (19-25, 26-35 and 36-45) from rush hours in the afternoon throughout to midnight.

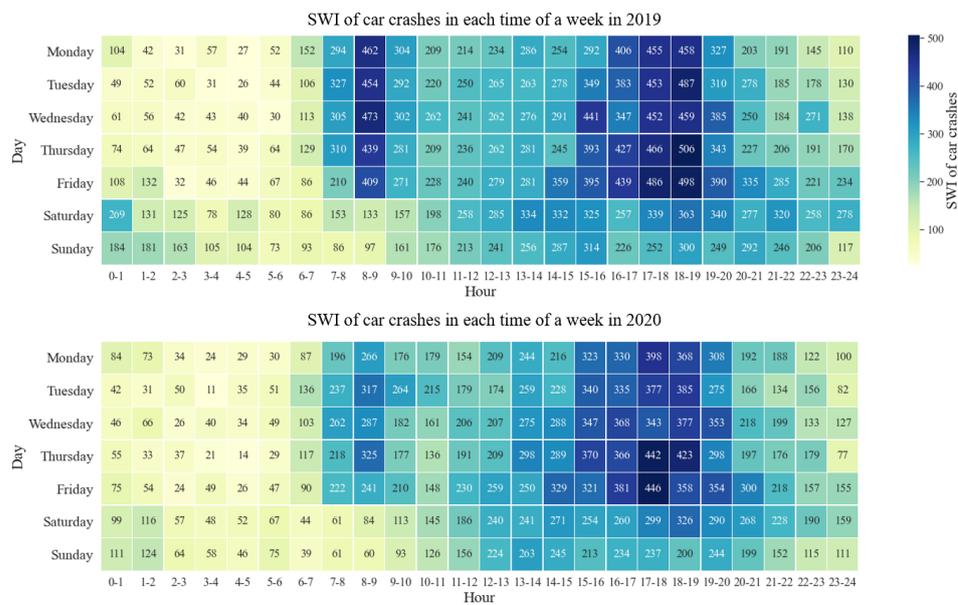

(a)

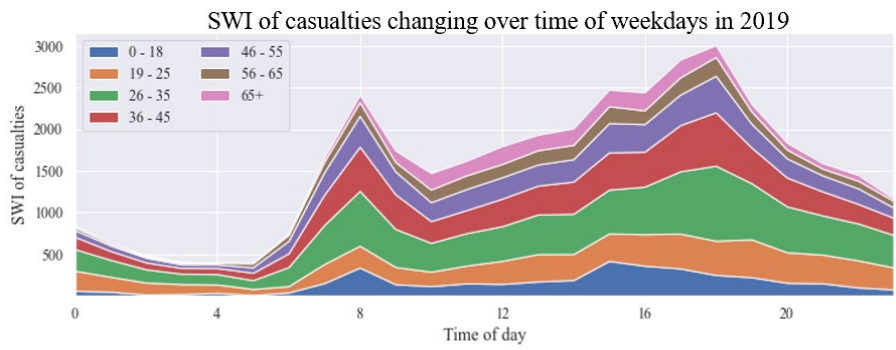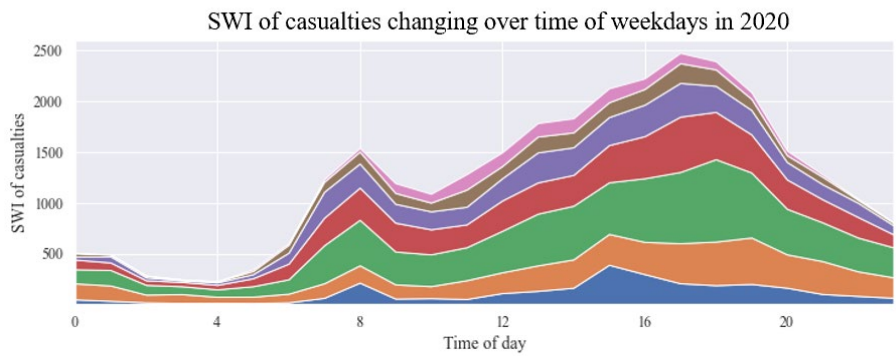

(b)

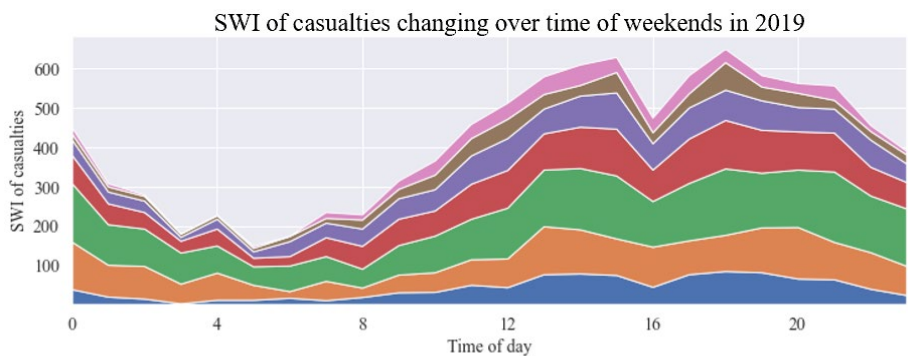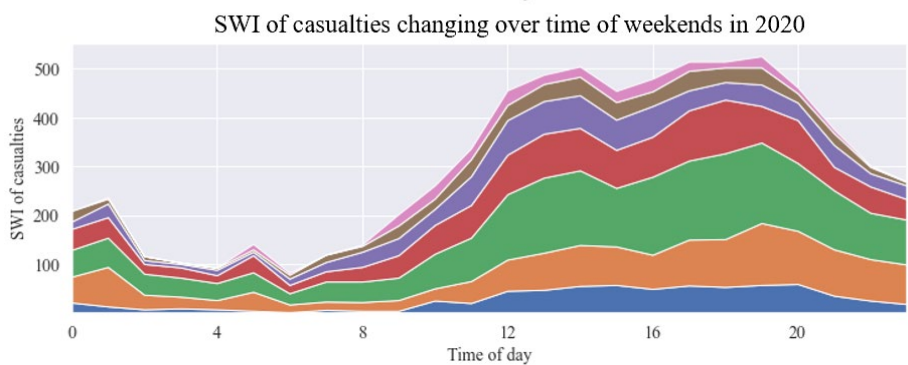

(c)

Figure 1. Temporal patterns of car crashes. (a) SWI of traffic crashes in each time of a week in 2019 & 2020; (b) SWI index of casualties changing over time of weekdays in 2019 & 2020; (c) SWI index of casualties changing over time of weekends in 2019 & 2020

### 4.2 Spatial analysis of road traffic crashes

High-risk locations of traffic crash in London pre- and para- the pandemic are detected by NetKDE, as shown in Figure 2, and colored in red. It illustrates that crashes are mainly distributed in the central part of London and extended in all directions alongside the road network, with most being located at road intersections and complex road structures such as roundabouts. Besides, there are smaller number of high-risk locations of crashes in 2020 than those in 2019, which shows a hollow distribution.

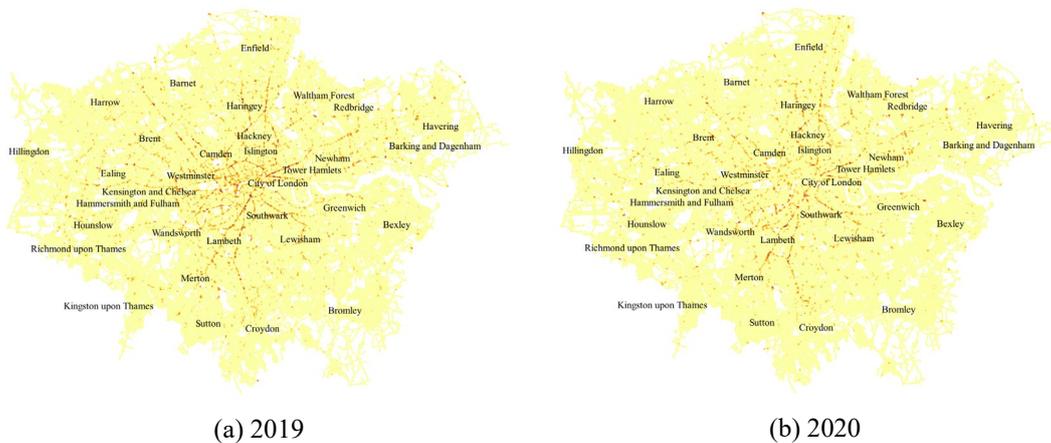

(a) 2019          (b) 2020

Figure 2. Network KDE of crashes in 2019 & 2020

To further explore the riskiest roads, the top four and first densest road segments in 2019 and 2020 respectively, which are in the same density interval, had been selected in North London or nearer to central London (Figure 3(a) & (b)). In 2019, the first road segment is Seven Sister Roads in the borough of Haringey of north London, in a busy multicultural neighbourhood near Finsbury Park and close to the major transport hub, the Finsbury Park Station featured busy local amenities. The second road is part of Regent Street in the borough of Westminster, hosting the most famous shopping centre and many flagship retail stores, shops, and restaurants. The third road locates at the intersection of Denmark Hill and Camberwell Green in Southwark, as the access road to one of the busiest stations in South London annoyed by congestion and overcrowding. The last road is Brixton Road in the borough of Lambeth, characterised by its being overcrowded but narrow with a pair of Bus Lines, and embracing several markets as well as the Brixton Tube Station. In 2020, there is only one riskiest road, Clapham High Street, which is near Brixton Road and has a similar characteristic. In terms of the hotspot analysis, it is clear that the pandemic and its prevention policies like lockdown have a great influence on crash hotspots.

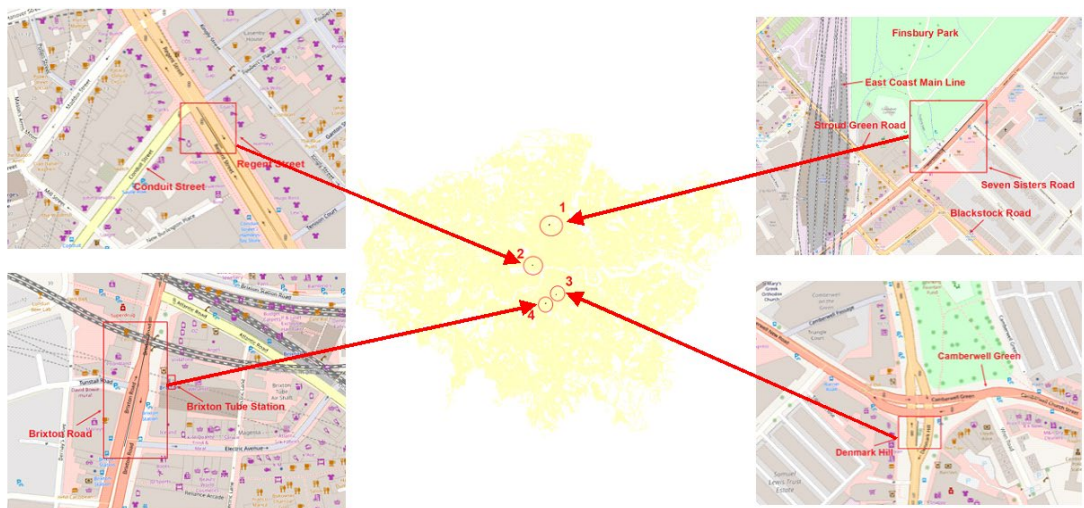

(a) 2019

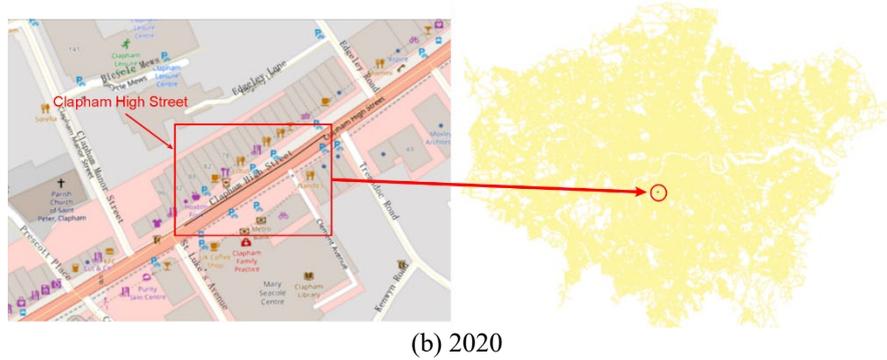

(b) 2020

Figure 3. The riskiest roads of Network KDE in 2019 & 2020

It is obvious that all these roads are bustling with POIs, i.e., shops, transport stations, and restaurants, evidencing the necessity to explore traffic crashes by varied POI categories.

The spatial cluster characteristics of two years' traffic crashes had been measured by Moran's *I* statistics, arriving at the significant global Moran's *I* at 0.25 and 0.18 (*p*-value of 0.001), respectively, indicating positive spatial autocorrelation hence clustering tendency at LSOA level of severity-weighted crashes. To further locate where are the clusters, LISA cluster map had been drawn using local Moran's *I* in Figure 4 (a) & (b). It could be found that the four most significant high-high (HH) clusters are in northern London, the city centre (high density of POIs), and on the west and east (the Heathrow airport location) wings of the city. In addition, Figure 4 (c) & (d) highlighted similarities and differences pre- and para- the pandemic, where the area in a black circle indicates those unchanged clusters, and areas in green or red circle are clusters experiencing an increase or decrease, respectively. For example, there are similar clusters of the Heathrow airport, Ealing borough (major metropolitan centre), and the City of London in 2019 and 2020, indicating the modest effect of the pandemic on such transport junctions and city centre. Majority of workplace clusters located in eastern London alongside the River Thames had shrunk in 2020, due to the work-from-home policy. On the contrary, parks and entertainments such as, Richmond Park, Hampton Court Park, Walthamstow Wetlands, etc., circled in green had expanded during the pandemics, showcasing citizens' rising leisure activities.

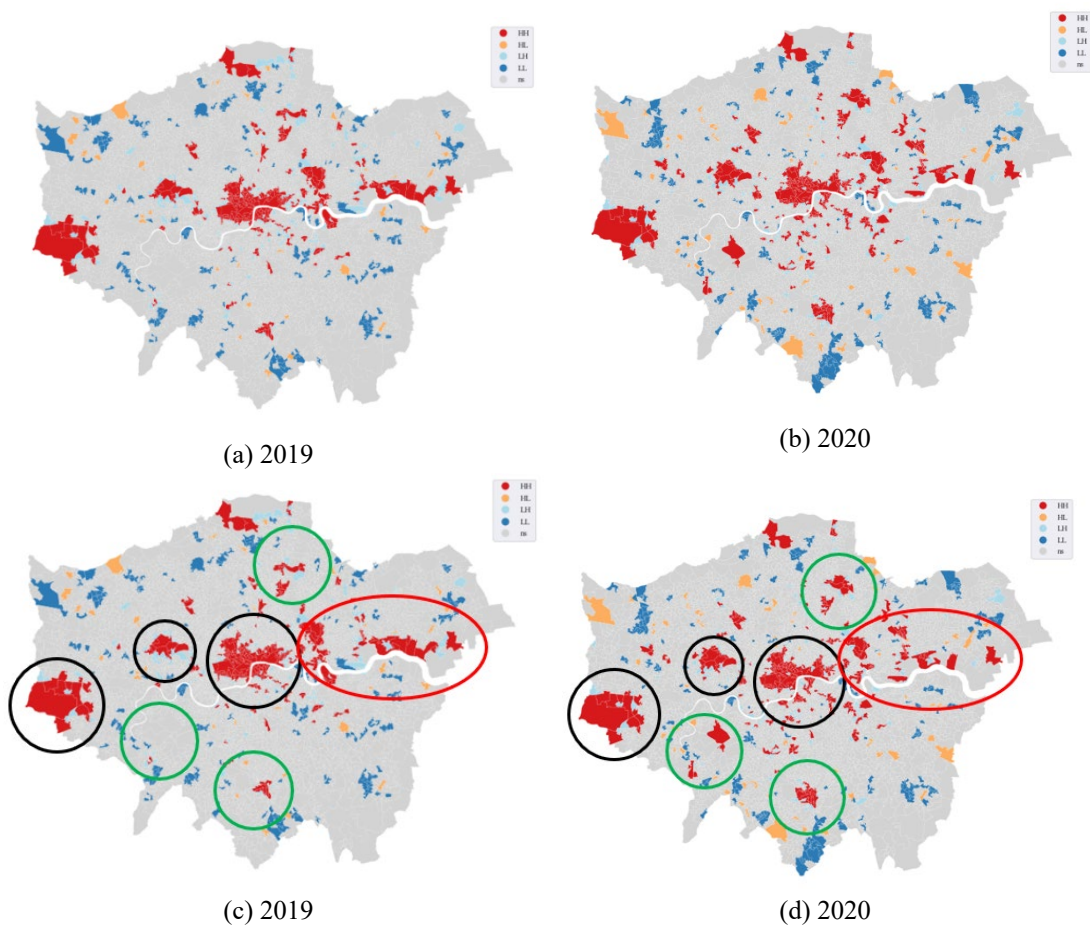

(a) 2019  (b) 2020

(c) 2019  (d) 2020

Figure 4. LISA cluster map for crashes in 2019 & 2020

## 4.3 Factor analysis of road traffic crashes

A geographical detector measures the individual and interactive effects of POIs on crash distribution and realise the POIs' categorisation through the JNB division method. A further factor detector implied the categorisation's statistical significance pre- and para- the pandemic, as shown in Table 4. In general, cliff drops of PD values had been witnessed but with unchanged rankings throughout the observation years, except for the rank for SHOP (🛒) increased from 5th to 3rd in 2020. In particular, the PD value for STATION (🚇) is the highest, with 43.2% and 29.6% of the spatial pattern of car crashes, being consistent with that of transport stations in 2019 and 2020, making it the dominant factor of crashes distribution. Besides, the PD values for WORK (💼) are second highest at 0.397 and 0.229 in these two years, respectively.

An interaction detector had been utilised to further quantify the interconnected effect of paired factors on crashes in 2019 and 2020. In theory, it would be 36 pairs of interactions among nine factors, but only listed the top five in Table 4 for information, with STATION as the compulsory. There are also declines of PD values for the consistent top four pairs of interactive factors, where the SHOP interacting with STATION that ranks from 3rd to 1st pre- and para- the pandemic. Another difference is that the 5th interactive factor have been changed from STATION interacting with ENTERTAINMENT (▶️) to its being interacting with EDUCATION (🏛️). As illustrated, the PD value of FOOD (🍴) interacting with STATION is the highest to explain 56.2% in 2019 and 36.5% in 2020 of the spatial patterns of crashes. Such results imply a possible high risk of car crashes in areas with a higher density of top five POIs near transport stations. In comparison with the individual effect, SHOP interacting with STATION para- and pre- the pandemic has the largest compound effect on crashes distribution, implying the significant altering and compounding effect for shops nearer transport stations onto car crashes. In addition, the climbing rankings of SHOP's PD values indicated one of the main daily activities for citizens during the pandemic was to go shopping rather than go out dining.

Table 2. PD values of individual and interactive effects of POIs in 2019 & 2020

| Effect type | POI categories | PD value(ranking) 2019 | PD value(ranking) 2020 | p-value |
|---|---|---|---|---|
| Individual effect |  | 0.432(1) | 0.296(1) | <0.05 |
| |  | 0.397(2) | 0.229(2) | <0.05 |
| |  | 0.366(3) | 0.200(4) | <0.05 |
| |  | 0.313(4) | 0.177(5) | <0.05 |
| |  | 0.296(5) | 0.207(3) | <0.05 |
| |  | 0.289(6) | 0.141(6) | <0.05 |
| |  | 0.220(7) | 0.121(7) | <0.05 |
| |  | 0.125(8) | 0.076(9) | <0.05 |
| |  | 0.088(9) | 0.089(8) | <0.05 |
| Interactive effect |  ∩  | 0.562(1) | 0.365(2) | <0.05 |
| |  ∩  | 0.554(2) | 0.363(3) | <0.05 |
| |  ∩  | 0.540(3) | 0.376(1) | <0.05 |
| |  ∩  | 0.510(4) | 0.344(4) | <0.05 |
| |  ∩  | 0.503(5) | - | <0.05 |
| |  ∩  | - | 0.323(5) | <0.05 |

## 4.4 Spatiotemporal data mining of road traffic crashes

To locate the locations and periods with higher car crashes rates by age group, this section models the crashes in each LSOA using two tensors, $\mathcal{A}_{weekday}, \mathcal{A}_{weekend} \in \mathbb{R}^{I \times J \times K}$, or weekdays and weekends in 2019 and 2020, respectively. The three dimensions in the tensors are $I$ for regions, $J$ for age groups, and $K$ for time slots in a single day. Since there are 4,835 LSOAs in London, $I$ will be presented by $r = [r_1, r_2, \cdots, r_{4835}]$; $J$ for age groups are $a = [a_1, a_2, \cdots, a_7]$ and $K$ for time slots by an hour as $t = [t_1, t_2, \cdots, t_{24}]$. The SWI index, therefore, is stored in entry $\mathcal{A}(i,j,k)$, which refers to a cumulative value of the SWI index of age groups $a_j$ in LSOA $r_i$ and time slot $t_k$ throughout the study period.

### *4.4.1 Tensor size selection*

The sizes of core tensors had been calculated respectively for 2019-weekday and 2020-weekday as well as 2019-weekend and 2020-weekend scenarios, with three decomposed time patterns for weekdays (morning peak hours, afternoon peak hours, and other times) and two distinctive patterns for weekends (evening peak hours and other times). After multiple attempts and integration of three similar age groups, the KL divergence parameter suggests a core tensor size at [13,5,3] for weekdays and [8,4,2] for weekends pre- and para- the pandemic (Figure 5). In Figure 5(a)&(c), the x-axis indicates different numbers of spatial patterns, and the y-axis lists the corresponding value of KL divergence, illustrating that the iteration of the parameter with initial value $KL = 3.1$ & $3.2$ converges to nearly 2.4 & 2.5 where $x = 13$ as the most appropriate size; the 2019-weekend's and 2020-weekend's selection in Figure 5(b)&(d) exhibits four main patterns by age group, with 8 optimal spatial patterns being suggested by the KL divergence.

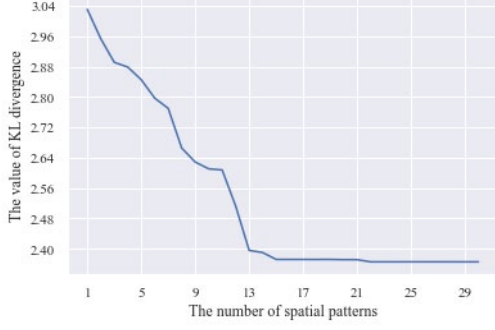 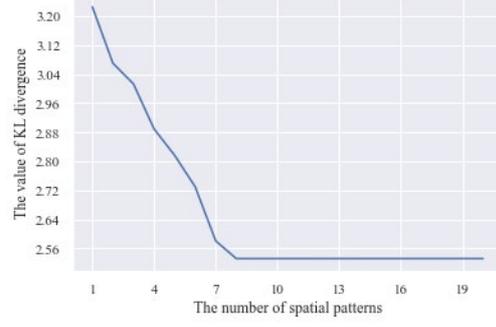

(a) KL divergence for 2019-weekday core tensor  (b) KL divergence for 2019-weekend core tensor

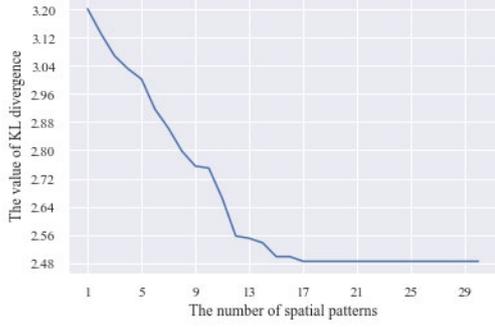 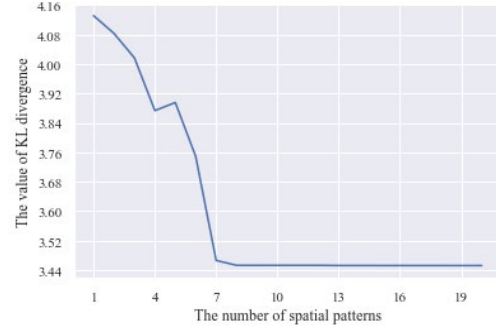

(c) KL divergence for 2020-weekday core tensor  (d) KL divergence for 2020-weekend core tensor

Figure 5. Size selections of four core tensors.

### 4.4.2 *Tensor decomposition*

These two tensors for car crashes in 2019 and 2020, respectively, $\mathcal{A}_{weekday}$ and $\mathcal{A}_{weekend}$, are decomposed separately into a core tensor and three matrices as presented: $\mathcal{G}_{weekday} \in \mathbb{R}^{13\times5\times3}$ and $\mathcal{G}_{weekend} \in \mathbb{R}^{8\times4\times2}$, Figure 6 shows the spatial patterns, age patterns, and temporal patterns on weekdays are identifiable as matrices of (a) $A_{weekday} \in \mathbf{R}^{4835\times13}$, $B_{weekday} \in \mathbf{A}^{7\times5}$, $C_{weekday} \in \mathbf{T}^{24\times3}$, and (b) $A_{weekend} \in \mathbf{R}^{4835\times8}$, $B_{weekend} \in \mathbf{A}^{7\times4}$, $C_{weekend} \in \mathbf{T}^{24\times2}$. It should be noted that the value of the SWI index storing each entry of tensors is normalised to [0,1] for nonnegative Tucker decomposition; the degree of importance for study subjects has been referred to as the "degree of importance", where a high value means a vital role of a target object in its pattern and vice versa.

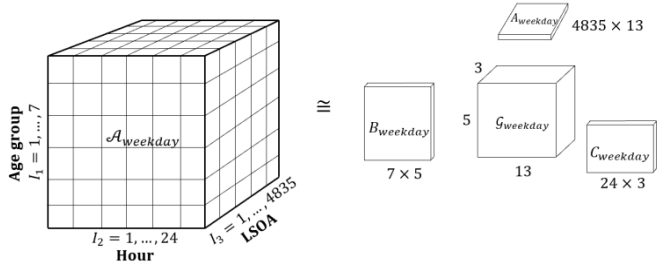

(a) 2019-weekday & 2020-weekday tensor

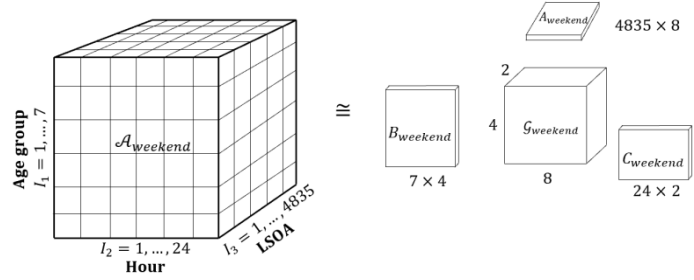

(b) 2019-weekend & 2020-weekend tensor

Figure 6. Tensor decomposition of tensors

### 4.4.3 Pattern analysis of road traffic crashes during weekdays

Upon tensor decomposition, Figure 7 depicted daily patterns, age patterns and a core tensor of traffic crashes on weekdays in 2019 and 2020, respectively. Figure 7(a) exhibited the obvious daily pattern pre- the pandemic in the green line, with a peak at 8:00 am, when car crashes occurred in the morning peak hours; this morning peak hour is not obvious in 2020 but the dinner peak hours with a peak at 5:00 pm. The orange line peaked at 5:00 pm - 6:00 pm in 2019 and 4:00 pm in 2020 are approximately the afternoon rush hours, implying the afternoon peaks of car crashes in 2020 appear earlier than those in 2019; the blue line illustrated the pattern during off-peak hours.

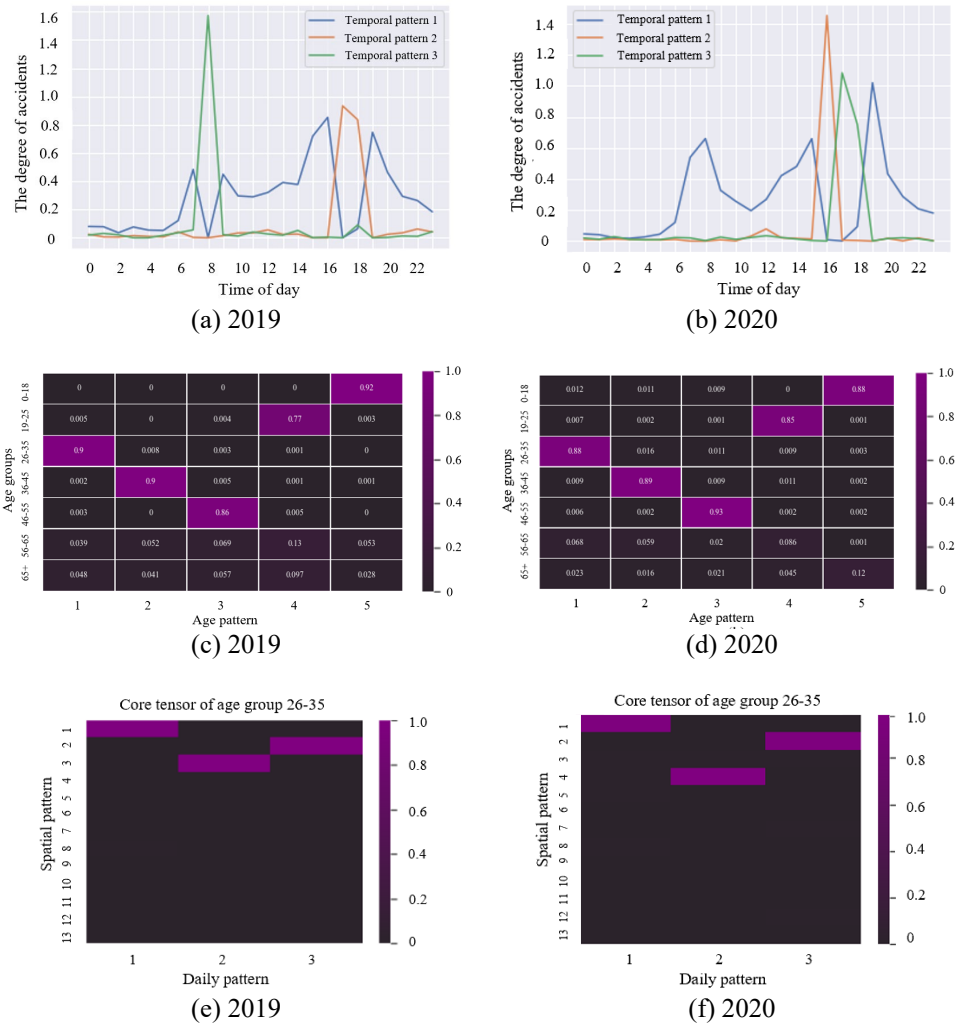

Figure 7. Decomposed patterns of 2019-weekday & 2020-weekday tensor. (a)&(b) Temporal pattern; (c)&(d) Age pattern; (e)&(f) Examples of matrices of Core tensor

*Note: the matrices of core tensor are only for illustrating our workflow but not for comparative analysis, so one example of this matrix is shown for two years respectively.*

In light of the varied age groups presented in Figure 7(c)&(d), the purple colour with higher proportions indicated the main subject of the age-group pattern, especially the patterns for 26–35-year-olds, 36–45-year-olds, 46–55-year-olds, 18–25-year-olds, and 0–18-year-olds. Other age group patterns were difficult to detect due to their possible high similarities. In total, 13 spatial patterns have been detected with complicated information about car crashes in these two years, contributing to the later geographical detector. To match with the core tensor reflected nexus of such patterns at three dimensions, four further sliced matrices from five age dimensions had been drawn in Figure 7(e)&(f), where the light purple indicates a high degree of closeness. For example, the first matrix reflected first age patterns for 26–35 age group, and it has 1$^{st}$ spatial pattern & off-peak hours daily pattern, 3$^{rd}$ spatial pattern & afternoon peak hour daily pattern, and 2$^{nd}$ spatial pattern & morning rush hour daily pattern. It should be noted that the third and fourth age patterns have a major pattern and minor pattern, rendering their potential similar patterns with other age groups.

### *4.4.4 Pattern analysis of road traffic crashes during weekends*

In a similar design, decomposed tensor results on weekends for two years were presented in Figure 8. As described in Section 4.1, most car crashes happened during the weekend evenings. In Figure 8(a), the orange line depicted that age-involved crashes occurred at 20:00 were distinctive, hence being reorganised as the pattern of evening peak hours; the blue line indicated car crashes distribution at other times on the weekends.

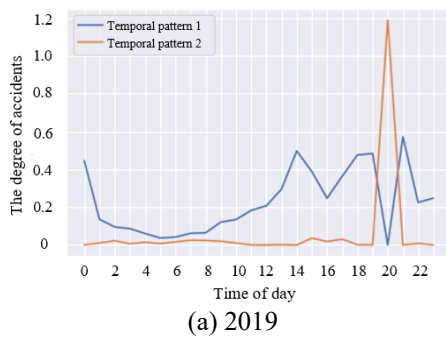
(a) 2019

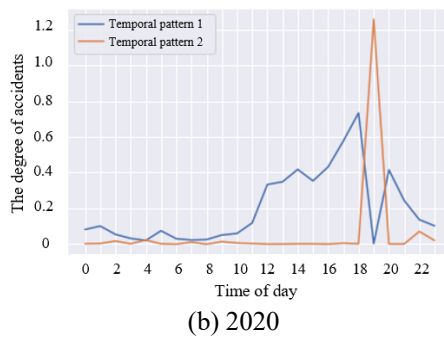
(b) 2020

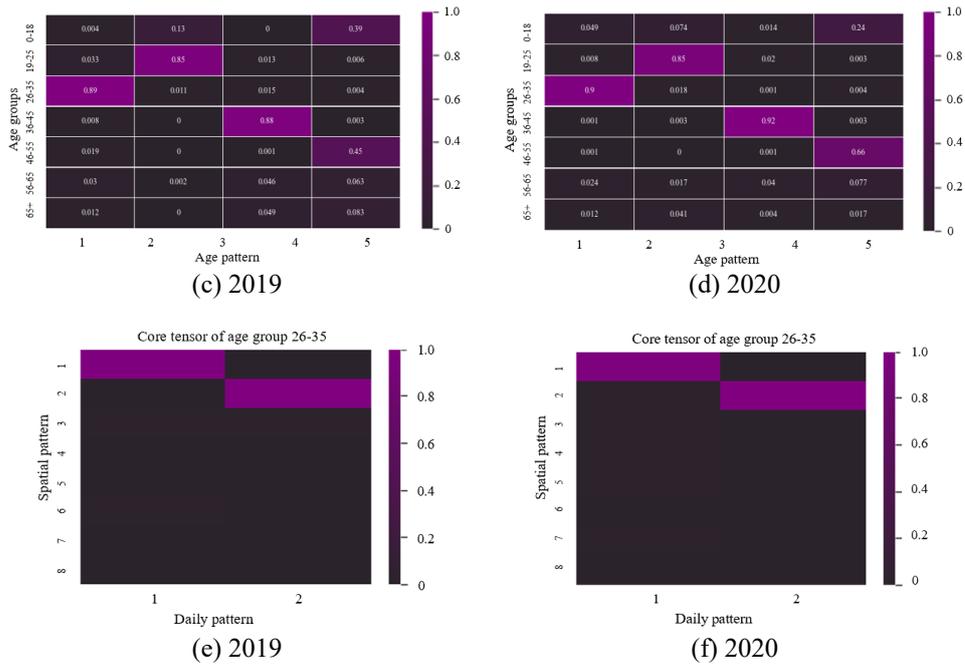

(c) 2019      (d) 2020

(e) 2019      (f) 2020

Figure 8. Decomposed patterns of weekday tensor. (a) Temporal pattern; (b) Age pattern; (c) Matrices of Core tensor

Figure 8(b)&(d) clearly exhibited four age patterns for people aged between 26 and 35, 18 to 25, 36 to 45, and 0-18 & 46-55. The patterns for people over 55 are not revealing due to their being similar as on weekdays. It could be observed that people aged 0-18 and 46-55 have the same crash patterns. There are eight spatial patterns and a core tensor for weekends being sliced by age dimensions but in identified metrics, with the last matrix representing the age patterns of 0-18 and 46-55.

*4.4.5 Spatiotemporal pattern analysis for each age group*

On basis of such spatial and temporal patterns in 2019 and 2020 by age group, both dominant POI factors and interactive effects of severity-weighted crashes could be explored through geographical detector, with results presented in Table 5 and Table 6.

According to Table 5, STATION and WORK factors have the same significant effects on car crashes involving every age group on weekdays pre- and para- the pandemic, except for victims aged above 55 years old. Nevertheless, it is obvious that every age group has different dominant factors. In terms of 0-18 age group specifically, POIs factors do not have significant PD-values in 2020, except for the EDUCATION () and its interaction with ENTERTAINMENT (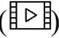) during dinner peak hour. It could be suggested that areas with a high density of schools, especially schools near ATTRACTIONS (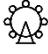), should be considered to protect the younger age group from crashes. Adults (age groups 19-25, 46-55), had similar changes of dominant factor, the SHOP () imposing its greater influence on patterns in 2020 than 2019, illustrating more car crashes near the areas with a high density of shops, especially SHOPS near ATTRACTIONS, ENTERTAINMENT places and WORKPLACES (), during afternoon and dinner peak hours. Besides, the pandemic also affects the factors for age group of 26-35 and 36-45. In particular, the RESTAURANT (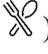) and WORK as well as their interactions with STATION () become the dominant factors for these two age groups in 2020, respectively, identifying those risky areas with a high density of WORKPLACES and RESTAURANTS around transport STATIONS.

Table 3. Dominant influence and interaction of factors for age groups during weekdays in 2019 & 2020

| Age group | Year | Off-peak hours Individual factor | Off-peak hours Interactive factors | Morning peak hours Individual factor | Morning peak hours Interactive factors | Afternoon peak hours Individual factor | Afternoon peak hours Interactive factors | Dinner peak hours Individual factor | Dinner peak hours Interactive factors |
|---|---|---|---|---|---|---|---|---|---|
| 0–18 | 19 | 🚇 | 🚇 ∩ 💼 | 🚇 | 🚇 ∩ ▶️ | - | - | 🚇 | 🚇 ∩ 💼 |
| 0–18 | 20 | - | - | - | - | - | - | 🏛️ | 🏛️ ∩ 🎡 |
| 19-25 | 19 | 🚇 | 🚇 ∩ 💼 | 🚇 | 🚇 ∩ 💼 | - | - | 🚇 | 🚇 ∩ ▶️ |
| 19-25 | 20 | 🚇 | 🚇 ∩ 💼 | - | - | 🛒 or 🚇 | 🛒 ∩ 🎡 | 💼 | 🛒 ∩ ▶️ |
| 26–35 | 19 | 🚇 | 🚇 ∩ 🍴 | 💼 | 🚇 ∩ 💼 | - | - | 🛒 | 🚇 ∩ 🛒 |
| 26–35 | 20 | 💼 | 🚇 ∩ 💼 | - | - | 🍴 | 🚇 ∩ 🍴 | 🍴 | 🍴 ∩ 🏛️ or 🚇 |
| 36–45 | 19 | 🚇 | 🚇 ∩ 🛒 | 🚇 | 🚇 ∩ 🏛️ | - | - | 🚇 | 🚇 ∩ 🏛️ |
| 36–45 | 20 | 🚇 | 🚇 ∩ 🛒 | - | - | 🍴 or 💼 | 🚇 ∩ 💼 | 🚇 | 🚇 ∩ 💼 |
| 46–55 | 19 | 🚇 | 🚇 ∩ 💼 | 🚇 | 🚇 ∩ 🍴 | - | - | 🚇 or 🍴 | 🚇 ∩ 🍴 |
| 46–55 | 20 | 🚇 | 🚇 ∩ 🎡 | - | - | 🚇 or 🛒 | 🛒 ∩ 💼 | 🚇 | 🚇 ∩ 🍴 |

In comparison to weekdays' factors, the WORK on weekends has lower effects on crash distribution and led out the dominant effects to ENTERTAINMENT and RESTAURANT. There are also differences of distinctive factors and interactions at the weekend by age group

pre- and para- the pandemic, especially during evening peak hours. For example, RESTAURANT becomes the dominant factor for age groups 0-18 and 46-55, during the evening peak hours, although STATION and its interactions with EDUCATION and ENTERTAINMENT have the dominant effect in 2020. Besides, during evening peak hours, the individual factors and interactive factors for 19-25 and 26-35 age groups have been changed from ENTERTAINMENT, RESTAURANT, and their interactions with STATION to PARKING, STATION, WORK and SHOP interacting with ENTERTAINMENT during the pandemic.

Table 4. Dominant influence and interaction of factors for age groups during weekends in 2019 & 2020

| Age group | Year | Off-peak hours | | Evening peak hours | |
|---|---|---|---|---|---|
| | | Individual factor | Interactive factors | Individual factor | Interactive factors |
| 0–18 | 19 | 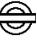 | 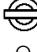 ∩ 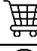 | 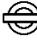 | 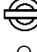 ∩ 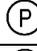 |
| | 20 | 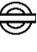 | 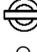 ∩ 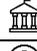 | 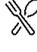 | 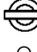 ∩ 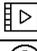 |
| 19-25 | 19 | 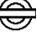 | 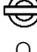 ∩ 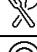 | 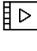 | 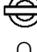 ∩ 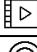 |
| | 20 | 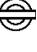 | 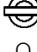 ∩ 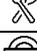 | 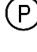 | 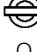 ∩ 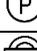 |
| 26–35 | 19 | 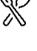 | 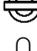 ∩ 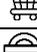 | 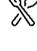 | 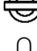 ∩ 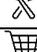 |
| | 20 | 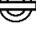 | 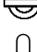 ∩ 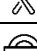 | 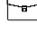 | 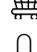 ∩ 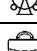 |
| 36–45 | 19 | 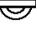 | 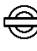 ∩ 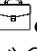 | 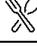 or 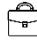 | |
| | | | 🍴 | | 🍴 |
| | 20 | 💼 | 🚇 ∩ 💼 | 💼 or 🍴 | 🚇 ∩ 🍴 |
| 46–55 | 19 | 🚇 | 🚇 ∩ 🛒 | 🚇 | 🚇 ∩ Ⓟ |
| | 20 | 🚇 | 🚇 ∩ 🏛 | 🍴 | 🚇 ∩ ▶ |

In summary, in the observing years, the distributions of transport stations, workplaces, and restaurants significantly affected casualty patterns of all age groups, while transport stations nearer to restaurants and workplaces have the largest impact on crashes patterns, calling for extra attention to deploy corresponding countermeasures in such areas to reduce traffic crashes risk behaviours, and to apply certain technique support in consolidating motorway safety engineering. Additionally, these results can be applied to identify the riskier areas requiring extra attention towards better traffic management and control strategies, tailoring for featured age groups through their featured activities pre- and para- the pandemic.

## 5 Conclusion

This study undertook an in-depth analysis of spatiotemporal patterns of crashes considering casualties by age group and injury severity, with the aim at addressing research questions on: "*What are the spatial and temporal patterns of car crashes as well as their changes in two typical years, 2019 and 2020, in London, and how do the influential factors work?*" and "*What *are the spatiotemporal patterns of casualty by age groups, and how people's daily activities affect the patterns before and during the pandemic*"?

Upon recognising the spatial and temporal variations of traffic crashes in London in two representative years towards recovery from pandemic, it had been found that weekdays and weekends exhibited distinct patterns with daily trends against the time of day. NetKDE technique had been utilised to detect the riskiest roads and identified the main high-risk locations were spatially distributed in central London but spread out along the road, with a higher likelihood to happen at busier places. Besides, the distribution of collisions was highly correlated to their neighbouring areas. Geographical detector had been utilised to quantify the influences from selected factors and their interactions, implying that areas with many transport stations and surrounding shops require particular attention to local transportation planning and traffic management. To further investigate the driving factors for crashes, the casualty patterns by age group had been detected from spatial and temporal dimensions, using tensor decomposition techniques. Hereby the study contributed to the literature on crashes' casualty and daily human activity analysis through its utilisation of geographical detector, in the results to discover the latent mechanism of spatiotemporal patterns for each group with defined POI categories. It was obvious that casualties on weekdays were more frequent than those on weekends, while the influential factors and interactions varied.

In a nutshell, detecting and understanding crash patterns before and during the pandemic can be of help to formulate deployment plans towards COVID-19 in varied age groups. It also found that these patterns and their factors have varied before and during the pandemic. During the pandemic, the traffic demand for public transport dramatically reduced, thanks to massive people's work from home, resulting in the cliff drop of crashes. However, it has been taken as a temporary phenomenon with a speedy recovery rise after a short-term decline. Such analyses in this study explore the spatial and temporal characteristics (Sekadakis et al., 2021) and factors that can support traffic management decision-making in the long run. Despite the contribution of this paper, some limitations could get improved in future extended work. For example, the spatiotemporal patterns of casualties aged over 55 have been unclear, which may owe to its relatively lower proportion of casualties hindered the representative pattern emerging; some more POI categories, such as hospitals and petrol stations, deserve to be added for comprehensive studies in the future towards a more in-depth study.

In the future, retrospective research on deploying data in 2021 will be conducted once with official datasets in place, to test the methodologies proposed in this study onto the real data and try to capture the full picture on London's road safety upon recovery from the pandemic.